\documentclass{emulateapj}
\usepackage{apjfonts}

\newcommand{\Msun}      {\mbox{$\rm\,M_{\mathord\odot}$}}

\begin{document}

\lefthead{The Be X-ray binary IGR J01363+6610 in quiescence}
\righthead{Tomsick et al.}

\submitted{To appear in the Astrophysical Journal}

\def\lsim{\mathrel{\lower .85ex\hbox{\rlap{$\sim$}\raise
.95ex\hbox{$<$} }}}
\def\gsim{\mathrel{\lower .80ex\hbox{\rlap{$\sim$}\raise
.90ex\hbox{$>$} }}}

\title{Confirmation of IGR J01363+6610 as a Be X-ray binary with
very low quiescent X-ray luminosity}

\author{John A. Tomsick\altaffilmark{1},
Craig Heinke\altaffilmark{2},
Jules Halpern\altaffilmark{3},
Philip Kaaret\altaffilmark{4},
Sylvain Chaty\altaffilmark{5},\\
Jerome Rodriguez\altaffilmark{5}, and
Arash Bodaghee\altaffilmark{1}}

\altaffiltext{1}{Space Sciences Laboratory, 7 Gauss Way, 
University of California, Berkeley, CA 94720-7450, USA
(e-mail: jtomsick@ssl.berkeley.edu)}

\altaffiltext{2}{Department of Physics, University of Alberta, 
Room 238 CEB, Edmonton, AB T6G 2G7, Canada}

\altaffiltext{3}{Columbia Astrophysics Laboratory, Columbia University, 
550 West 120th Street, New York, NY 10027-6601, USA}

\altaffiltext{4}{Department of Physics and Astronomy, University of
Iowa, Iowa City, IA 52242, USA}

\altaffiltext{5}{AIM - Astrophysique Instrumentation Mod\'elisation
(UMR 7158 CEA/CNRS/Universit\'e Paris 7 Denis Diderot),
CEA Saclay, DSM/IRFU/Service d'Astrophysique, B\^at. 709,
L'Orme des Merisiers, FR-91 191 Gif-sur-Yvette Cedex, France}

\begin{abstract}

The field containing the candidate High Mass X-ray Binary IGR~J01363+6610 
was observed by {\em XMM-Newton} on 2009 July 31 for 28~ks.  A Be star 
was previously suggested as the possible counterpart of the {\em INTEGRAL} 
source, and although {\em Chandra}, during a 2007 observation, did not 
detect an X-ray source at the position of the Be star, we find a variable
source (XMMU J013549.5+661243) with an average X-ray flux of $2\times 10^{-13}$ 
ergs~cm$^{-2}$~s$^{-1}$ (0.2--12~keV, unabsorbed) at this position with 
{\em XMM-Newton}.  The spectrum of this source is consistent with a
hard power-law with a photon index of $\Gamma = 1.4\pm 0.3$ and a
column density of $N_{\rm H} = (1.5^{+0.7}_{-0.5})\times 10^{22}$ cm$^{-2}$
(90\% confidence errors).  These results, along with our optical
investigation of other X-ray sources in the field, makes the association
with the Be star very likely, and the 2~kpc distance estimate for the Be 
star indicates an X-ray luminosity of $9.1\times 10^{31}$ ergs~s$^{-1}$.  
This is lower than typical for a Be X-ray binary, and the upper limit on 
the luminosity was even lower ($<$$1.4\times 10^{31}$ ergs~s$^{-1}$ assuming 
the same spectral model) during the {\em Chandra} observation.  We discuss
possible implications of the very low quiescent luminosity for the 
physical properties of IGR~J01363+6610.

\end{abstract}


\section{Introduction}

The hard X-ray imaging of the Galactic plane by the {\em International 
Gamma-Ray Astrophysics Laboratory (INTEGRAL)} satellite \citep{winkler03} 
has uncovered a large number of new or previously poorly studied ``IGR'' 
sources \citep{bodaghee07,bird10}.  While {\em INTEGRAL} excels at detecting 
sources in the 20--50~keV band, it only localizes the sources to 
$1^{\prime}$--$5^{\prime}$, requiring follow-up observations with other X-ray 
satellites to obtain secure optical or IR counterparts, allowing for a 
determination of the nature of the sources \citep{walter06,tomsick08a,rtc09}.

As more and more of these sources have been identified, possibly the biggest 
surprise is the large number of High Mass X-ray Binaries (HMXBs) as well as 
the properties of these systems.  Many of the dozens of {\em INTEGRAL} HMXBs 
\citep{bodaghee07} have large levels of intrinsic absorption with 
$N_{\rm H}\sim 10^{23}$--$10^{24}$ cm$^{-2}$ \citep[e.g.,][]{walter06}, and these 
are commonly called obscured HMXBs.  In many cases, it appears that this is 
due to the compact object being embedded in a strong stellar wind 
\citep{fc04,moon07,chaty08}.  Some members of the group of {\em INTEGRAL} 
HMXBs exhibit other extreme properties, including the high amplitude X-ray 
flaring of the Supergiant Fast X-ray Transients \citep[SFXTs,][]{intzand05,smith06} 
or long-period pulsations from very slowly rotating neutron stars \citep{patel07}.

{\em INTEGRAL} observations have also led to the addition of more HMXBs in
the Be X-ray binary class \citep{rv82}.  The optical flux from these systems 
is dominated by an early-type star with emission lines from a circumstellar
disk.  In most cases, transient X-ray emission demonstrates the binary nature
of the system as eccentric orbits lead to periodic X-ray outbursts when the
compact object approaches periastron.  Of the 64 known Be X-ray binary systems, 
X-ray pulsations indicate the presence of a neutron star in 42 cases, and the 
compact object type is unknown for the remaining systems \citep{bz09}.  One of 
the interesting properties of this class is the observed correlation between 
the orbital period and the spin period of the neutron star \citep{corbet86}. 

IGR~J01363+6610 has been tentatively classified as a Be X-ray binary.  The 
source was discovered during {\em INTEGRAL} observations on 2004 April 19 
but was not detected $\sim$2 weeks later, indicating that the source is 
transient \citep{grebenev04}.  The peak flux observed from the source was 
17 mcrab ($\sim$$2.6\times 10^{-10}$ ergs~cm$^{-2}$~s$^{-1}$) in the 17--45~keV 
band and 9 mcrab ($\sim$$9.1\times 10^{-11}$ ergs~cm$^{-2}$~s$^{-1}$)
in the 8--15~keV band \citep{grebenev04}.  The 3-$\sigma$ upper limit in
the 17--45~keV band 2 weeks later was $<$11 mcrab \citep{grebenev04}, and
the source has not been detected in other {\em INTEGRAL} observations
even though 2.3~Ms of {\em INTEGRAL} time have been accumulated at the
position of this source \citep{bird10}.  Reports of non-detection include
both detailed studies of the Cassiopeia region \citep{denhartog06} 
and catalogs indicating that the source was only detected during the
discovery outburst in 2004 \citep{krivonos07,bird07,bird10}.

Although the {\em INTEGRAL} position uncertainty of $3^{\prime}\!.7$ leaves 
a large error region, a Be star was found within the error circle using 
narrow-band H$\alpha$ imaging and follow-up optical spectroscopy, and it
has been suggested to be the likely counterpart \citep{reig04a,reig05}.  
However, a sensitive X-ray observation taken with the {\em Chandra X-ray 
Observatory} in 2007 failed to detect the Be star \citep{tomsick08a}.
At the 2~kpc distance estimated for the Be star \citep{reig05}, the 
non-detection implies an upper limit on the X-ray luminosity of
$<$$2\times 10^{31}$ ergs~cm$^{-2}$~s$^{-1}$ \citep{tomsick08a}, which is 
lower than quiescent luminosities for other Be X-ray binaries 
\citep{campana02a}.  This luminosity approaches the level that has
been seen during quiescent periods from transient Low Mass X-ray Binaries 
(LMXBs) due to thermal emission from the neutron star surface 
\citep{bbr98,campana98}.  Although neutron stars in HMXBs might not be 
heated to the high levels seen for LMXBs, it is interesting that these 
observations probe this luminosity regime.

Currently, there are significant uncertainties about the nature of 
IGR~J01363+6610.  While we know that it is an X-ray transient, and the
hard X-ray emission makes it likely that it is a binary (although its
orbital period is unknown), the {\em Chandra} non-detection makes it 
unclear whether it is really a Be X-ray binary.  Finding a Be star in 
the relatively large {\em INTEGRAL} error circle is not convincing 
because Be stars are more commonly found as single stars rather than 
being part of a binary system \citep{pr03}.  Furthermore, we cannot be 
certain that the compact object in the system is a neutron star since 
pulsations have not been detected.

In this paper, we report on a second sensitive X-ray observation of the 
IGR~J01363+6610 field with {\em XMM-Newton} along with optical spectroscopy
of the Be star as well as other X-ray sources with optical counterparts 
in the field.  With {\em XMM-Newton}, we confirm that the Be star is an
X-ray source.  We also re-analyze the {\em Chandra} observation and discuss 
the 2007 results in the context of the new information from {\em XMM-Newton}.

\section{Observations}

\subsection{XMM-Newton}

We observed the IGR~J01363+6610 field with {\em XMM-Newton} on 2009 
July 31 from 14.4~h to 22.3~h UT.  The observation (ObsID 0603850101)
occurred during {\em XMM-Newton} revolution 1766.  The EPIC pn, MOS1, 
and MOS2 instruments \citep{struder01,turner01} were all operated in 
Full Frame mode with a medium filter.  We used the {\em XMM-Newton} 
Science Analysis Software (SAS-10.0.0) package to process the raw data 
files.  We used the SAS tool {\ttfamily cifbuild} to obtain the necessary 
``current calibration files'' for the observation.  Then, we reprocessed 
the pn, MOS1, and MOS2 data using {\ttfamily epproc} and {\ttfamily emproc}, 
yielding photon event lists for the three instruments.  Although proton 
flares sometimes cause portions of {\em XMM-Newton} observations to have 
very high backgrounds, we did not find evidence for proton flares, and 
we were able to use the full exposure time.

\subsection{Chandra X-ray Observatory}

We also used an observation of the IGR~J01363+6610 field made by the
{\em Chandra X-ray Observatory}.  The observation (ObsID 7533), which
had an exposure time of 4,976~s, was made on 2007 June 8.  We used the
Advanced CCD Imaging Spectrometer \citep[ACIS;][]{garmire03}, and the
aimpoint was placed on the ACIS-S3 chip.  The 90\% confidence 
{\em INTEGRAL} error circle for IGR~J01363+6610 \citep{bird10} is
contained on the ACIS-S3 and ACIS-S4 chips.  Results of the observation
were previously reported in \cite{tomsick08a}, and the procedures used
for data reduction are described in that work.

\subsection{MDM Optical Observations}

Optical spectroscopic observations were made with the Boller \& Chivens
CCD spectrograph on the 2.4~m Hiltner Telescope of the MDM Observatory
on 2009 August 23 UT.  Conditions were clear, but with high humidity.
The 150 g/mm grating, blazed at 4700~\AA, was used with a 1$^{\prime\prime}$
wide slit, which gave a wavelength coverage of 3700$-$7360 \AA\ at 
7.6 \AA\ resolution.   Spectra of six stars coinciding with X-ray sources 
in the field of IGR~J01363+6610 were obtained, each with an exposure time 
of 120~s.  Of these, three are inside the refined {\it INTEGRAL\/} error 
circle and are presented here.  Spectral reduction was performed using 
standard IRAF procedures.  Flux calibration used \cite{og83} standard stars, 
although the spectrophotometry is not expected to be accurate because of 
the narrow slit.

\section{Results}

\subsection{X-ray Sources in the {\em INTEGRAL} Error Circle}

For the {\em XMM-Newton} analysis, we used the SAS tool {\ttfamily edetect\_chain} 
to search for sources in the $\sim$30$^{\prime}$ diameter field-of-view (FOV) of 
the MOS1 and MOS2 instruments, and included photons in the 0.1--10~keV energy band.  
All seven of the MOS2 CCDs were operational, but, for MOS1, CCD\#6 was not active 
because it was switched off after an anomaly that occurred in 2005.  While this 
means that some of the MOS2 FOV is not covered by MOS1, the central CCD, which 
includes the entire {\em INTEGRAL} error circle, is covered by both MOS units.

We found a total of 21 sources, including 14 detected by both MOS units and 
seven detected by MOS2 in the part of the FOV that MOS1 did not cover.  We 
determined the number of counts for each source using $25^{\prime\prime}$ radius 
apertures and subtracted the background using circular apertures of the same size 
in 16 source-free regions of the detector.  The mean number of background counts 
per aperture is $47.1\pm 1.8$ for MOS1 and $45.9\pm 1.7$ for MOS2 (in 28,180~s 
of exposure time).  The sources range in brightness from $12\pm 8$ to $156\pm 13$ 
counts (both of these sources are in the region covered only by MOS2).  Five of 
the sources are in the $3^{\prime}\!.7$ {\em INTEGRAL} 90\% confidence error circle 
for IGR~J01363+6610 given in \cite{bird10}.  These five sources include the second
brightest source in the field ($132\pm 9$ counts) as well as sources with 
$74\pm 8$, $31\pm 6$, $25\pm 6$, and $19\pm 6$ counts, where these numbers of
counts are averages of the two MOS detectors.

The {\em XMM-Newton} source names and positions are given in Table~\ref{tab:xmmsources}. 
In addition to MOS1 and MOS2, we also determined the position measured by the pn
instrument for each source, and we report the average position measured by the
three instruments (except for XMMU~J013632.4+660924, which fell between two pn 
CCD chips).  In averaging the positions, we weighted each measurement by its 
statistical uncertainty.  In Table~\ref{tab:xmmsources}, we also report the 
overall 90\% confidence position uncertainties, and, in each case, the error
is dominated by the systematic pointing uncertainty of $3^{\prime\prime}\!.4$\footnote{In
the document entitled ``EPIC status of calibration and data analysis'' 
(XMM-SOC-CAL-TN-0018), Guainazzi et al.~report an rms value for the systematic
pointing uncertainty of $2^{\prime\prime}\!.0$, and we have multiplied this by
1.7 to obtain the 90\% confidence value.}

\begin{table*}
\caption{{\em XMM-Newton} Sources in the {\em INTEGRAL} Error Circle\label{tab:xmmsources}}
\begin{minipage}{\linewidth}
\begin{center}
\footnotesize
\begin{tabular}{lcccc} \hline \hline
Name & R.A. (J2000)\footnote{The position is the weighted average of the positions measured by the three {\em XMM-Newton} instruments.} & Decl. (J2000)$^{a}$ & Position & MOS\\
     &              &               & Uncertainty\footnote{The 90\% confidence uncertainty in the {\em XMM-Newton} position.  This includes a systematic contribution of $3^{\prime\prime}.4$ due to the absolute pointing uncertainty and a statistical contribution.  We have added the two contributions in quadrature.} & counts\footnote{The average of the 0.1--10~keV count rates measured by MOS1 and MOS2.}\\ \hline
XMMU J013549.5+661243 & $01^{\rm h}35^{\rm m}49^{\rm s}.53$ & +$66^{\circ}12^{\prime}43^{\prime\prime}\!.1$ & $\pm$$3^{\prime\prime}\!.4$ & $132\pm 9$\\
XMMU J013606.5+661304 & $01^{\rm h}36^{\rm m}06^{\rm s}.54$ & +$66^{\circ}13^{\prime}04^{\prime\prime}\!.5$ & $\pm$$3^{\prime\prime}\!.9$ & $25\pm 6$\\
XMMU J013620.8+660851 & $01^{\rm h}36^{\rm m}20^{\rm s}.80$ & +$66^{\circ}08^{\prime}51^{\prime\prime}\!.0$ & $\pm$$3^{\prime\prime}\!.7$ & $31\pm 6$\\ 
XMMU J013632.4+660924\footnote{This source is detected by MOS1 and MOS2, but not by the pn instrument.  It falls on a gap between pn CCD chips.} & $01^{\rm h}36^{\rm m}32^{\rm s}.48$ & +$66^{\circ}09^{\prime}24^{\prime\prime}\!.0$ & $\pm$$4^{\prime\prime}\!.1$ & $19\pm 6$\\
XMMU J013644.2+661302 & $01^{\rm h}36^{\rm m}44^{\rm s}.26$ & +$66^{\circ}13^{\prime}02^{\prime\prime}\!.2$ & $\pm$$3^{\prime\prime}\!.5$ & $74\pm 8$\\ \hline
\end{tabular}
\end{center}
\end{minipage}
\end{table*}

Using the 2007 observation of the field by {\em Chandra}, we previously 
reported three sources in the {\em INTEGRAL} error circle \citep{tomsick08a}.  
However, in the most recent {\em INTEGRAL} source catalog \citep{bird10}, 
the best estimate of the IGR~J01363+6610 position has shifted by $1^{\prime}.8$
relative to the value used in \cite{tomsick08a}, and a fourth {\em Chandra}
source, CXOU~J013644.5+661301, is now also consistent with the {\em INTEGRAL}
position for IGR~J01363+6610, and its {\em Chandra} position is  R.A.~(J2000) = 
$01^{\rm h}36^{\rm m}44^{\rm s}.54$, Decl.~(J2000) = +$66^{\circ}13^{\prime}01^{\prime\prime}\!.6$
with a 90\% confidence uncertainty of $0^{\prime\prime}\!.64$.  Of the four 
{\em Chandra} sources, two are coincident with {\em XMM-Newton} sources.  Thus, 
merging the 2007 {\em Chandra} source list with the list of 2009 {\em XMM-Newton} 
leaves a list of seven X-ray sources in the {\em INTEGRAL} error circle, and these 
sources are listed in Table~\ref{tab:sources}.

\begin{table*}
\caption{Optical/Infrared Identifications\label{tab:sources}}
\begin{minipage}{\linewidth}
\begin{center}
\footnotesize
\begin{tabular}{ccccc} \hline \hline
Catalog/Source\footnote{The catalogs are the 2 Micron All-Sky Survey (2MASS) and
the United States Naval Observatory (USNO-B1.0).} & Separation\footnote{The uncertainty on the separation is the addition (in quadrature) of the X-ray source position error and the 2MASS or USNO-B1.0 error.} & & Magnitudes & \\ \hline
\multicolumn{5}{c}{(1) XMMU J013549.5+661243}\\ \hline
USNO-B1.0 1562-0030282 & $2^{\prime\prime}\!.0\pm 3^{\prime\prime}\!.4$ & $B = 14.9\pm 0.3$ & $R = 12.4\pm 0.3$ & $I = 11.1\pm 0.3$\\
2MASS J01354986+6612433 & $2^{\prime\prime}\!.0\pm 3^{\prime\prime}\!.4$ & $J = 10.04\pm 0.02$ & $H = 9.57\pm 0.03$ & $K_{s} = 9.12\pm 0.02$\\ \hline
\multicolumn{5}{c}{(2) XMMU J013606.5+661304}\\ \hline
USNO-B1.0 1562-0030364 & $2^{\prime\prime}\!.5\pm 3^{\prime\prime}\!.9$ & $B = 12.1\pm 0.3$ & $R = 10.7\pm 0.3$ & $I = 10.1\pm 0.3$\\
2MASS J01360684+6613021 & $3^{\prime\prime}\!.0\pm 3^{\prime\prime}\!.9$ & $J = 10.42\pm 0.02$ & $H = 10.10\pm 0.03$ & $K_{s} = 10.00\pm 0.02$\\ \hline
\multicolumn{5}{c}{(3) CXOU J013609.9+661157}\\ \hline
\multicolumn{5}{l}{The closest USNO-B1.0 source is $9^{\prime\prime}\!.1\pm 0^{\prime\prime}\!.6$ away.}\\
\multicolumn{5}{l}{The closest 2MASS source is $9^{\prime\prime}\!.0\pm 0^{\prime\prime}\!.6$ away.}\\ \hline
\multicolumn{5}{c}{(4) XMMU J013620.8+660851}\\ \hline
\multicolumn{5}{l}{The closest USNO-B1.0 source is $7^{\prime\prime}\!.2\pm 3^{\prime\prime}\!.7$ away.}\\
\multicolumn{5}{l}{The closest 2MASS source is $8^{\prime\prime}.\!7\pm 3^{\prime\prime}\!.7$ away.}\\ \hline
\multicolumn{5}{c}{(5) CXOU J013621.2+660928}\\ \hline
USNO-B1.0 1561-0031113  & $0^{\prime\prime}\!.52\pm 0^{\prime\prime}\!.64$ & $B = 17.2\pm 0.3$ & $R = 13.6\pm 0.3$ & $I = 12.0\pm 0.3$\\
2MASS J01362115+6609286 & $0^{\prime\prime}\!.38\pm 0^{\prime\prime}\!.64$ & $J = 10.42\pm 0.03$ & $H = 9.54\pm 0.03$ & $K_{s} = 9.26\pm 0.02$\\ \hline
\multicolumn{5}{c}{(6) XMMU J013632.4+660924/CXOU J013632.8+660924}\\ \hline
\multicolumn{5}{l}{The closest USNO-B1.0 source is $9^{\prime\prime}\!.1\pm 0^{\prime\prime}\!.6$ away.}\\
\multicolumn{5}{l}{The closest 2MASS source is $7^{\prime\prime}\!.5\pm 0^{\prime\prime}\!.6$ away.}\\ \hline
\multicolumn{5}{c}{(7) XMMU J013644.2+661302/CXOU J013644.5+661301}\\ \hline
USNO-B1.0 1562-0030589 & $0^{\prime\prime}\!.73\pm 0^{\prime\prime}\!.64$ & $B = 17.4\pm 0.3$ & $R = 14.4\pm 0.3$ & $I = 13.1\pm 0.3$\\
2MASS J01364458+6613014 & $0^{\prime\prime}\!.33\pm 0^{\prime\prime}\!.64$ & $J = 12.77\pm 0.02$ & $H = 12.01\pm 0.03$ & $K_{s} = 11.83\pm 0.03$\\ \hline
\end{tabular}
\end{center}
\end{minipage}
\end{table*}

\subsection{Optical and IR Counterparts and Optical Spectroscopy}

\begin{figure}
%\plotone{fig1.ps}
\includegraphics[clip,scale=0.45]{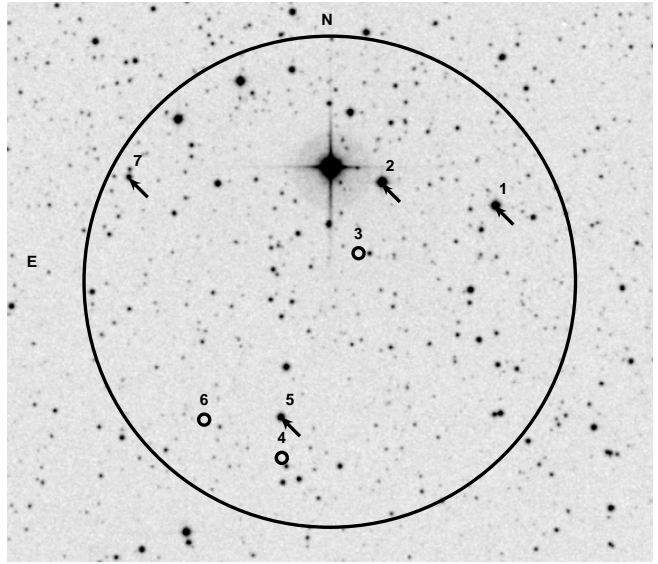}
\caption{Red optical image from the Digitized Sky Survey.  The large 
circle is the $3^{\prime}\!.7$ {\em INTEGRAL} error circle from 
\cite{bird10}.  The seven X-ray sources detected by {\em XMM-Newton}
and {\em Chandra} are labeled.  The arrows point to the optical 
counterparts for four of the X-ray sources, and their optical and IR 
magnitudes are given in Table~\ref{tab:sources}. The small circles
mark the locations of the X-ray sources without known optical 
counterparts.\label{fig:rimage}}
\end{figure}

For the seven X-ray sources listed in order of R.A. in Table~\ref{tab:sources}, 
we searched for optical and infrared (IR) counterparts in the United States Naval 
Observatory (USNO-B1.0) and 2 Micron All-Sky Survey (2MASS) catalogs.  Four of 
the sources (designated as \#1, \#2, \#5, and \#7) have optical and IR counterparts, 
and the names of these counterparts and their magnitudes are given in 
Table~\ref{tab:sources}.  For the other three X-ray sources (\#3, \#4, and \#6), 
the nearest USNO and 2MASS sources are $>$7$^{\prime\prime}$ away, which indicates 
that these X-ray sources are not associated with any sources in the USNO or 2MASS 
catalogs.  Figure~\ref{fig:rimage} shows a red (close to $R$-band) optical image 
from the Digitized Sky Survey with the revised {\em INTEGRAL} error circle
and the seven X-ray sources labeled.

\begin{figure}
\includegraphics[scale=0.33,angle=270]{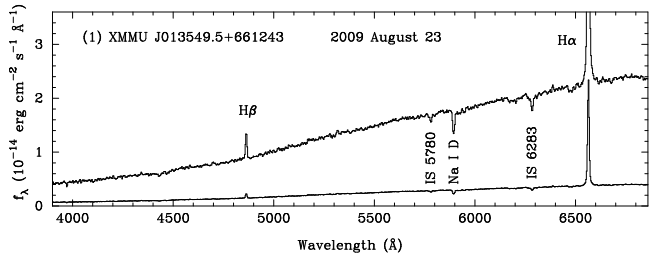}
\caption{Optical spectrum of the Be star XMMU~J013549.5+661243 taken on 
2009 August 23, which is a few weeks after the {\em XMM-Newton} observation.  
The optical spectrum was taken with the MDM~2.4 meter telescope and shows 
strong H$\alpha$ and H$\beta$ lines, indicating the presence of a 
circumstellar disk.  The lower spectrum is the upper spectrum divided by 6, 
allowing for all of the H$\alpha$ line to be visible.  This source was 
previously suggested as the most likely counterpart to IGR~J01363+6610, 
and the detection of the Be star with {\em XMM-Newton} provides further
confirmation of this association.\label{fig:optical}}
\end{figure}

X-ray source \#1 is the Be star that was previously identified using optical 
imaging and spectroscopy taken in 2004 \citep{reig04a,reig05}.  The optical spectrum 
of this star (USNO-B1.0~1562-0030282) is shown in Figure~\ref{fig:optical} and has 
the blue continuum and the H$\alpha$ and H$\beta$ emission lines indicative of a 
Be star.  The equivalent width (EW) of H$\alpha$ is --$54\pm 3$~\AA, which is consistent 
with the value measured by \cite{reig04a}, suggesting that the Be star's circumstellar
disk is stable.  Interstellar absorption features in the 
spectrum can be used to estimate the extinction $E(B-V)$ following the correlations 
in \cite{herbig75}.  Most commonly, the $4430$~\AA\ diffuse interstellar band is 
used for this purpose, but it falls in a poorly exposed region of our spectrum.
Instead we use the 5780~\AA\ feature.  With an EW of 1.0~\AA, it corresponds to 
$E(B-V)$ in the range 1.5$-$2.0, which is consistent with the estimate of 
\cite{reig05} based on the spectral classification and photometry, $E(B-V)=1.6$.

X-ray source \#2 is coincident with USNO-B1.0~1562-0030364, which is also known
as the bright ($V = 11.5$) optical source TYC~4043-860-1.  There is some 
confusion about whether this is an emission line star based on a catalog of
stars with H$\alpha$ in emission \citep{gg56}.  However, H$\alpha$ images and
optical spectra taken, respectively, 2 months and 5 months after the X-ray outburst 
from IGR~J01363+6610 do not show evidence for H$\alpha$ emission and the 
spectrum shows H$\alpha$ in absorption \citep{reig04a}.  The {\em XMM-Newton}
observation provides the first evidence that this is also an X-ray source, but
the lack of an H$\alpha$ emission line in 2004 makes it very unlikely to be the 
correct IGR~J01363+6610 counterpart.

\begin{figure}
\includegraphics[scale=0.33,angle=270]{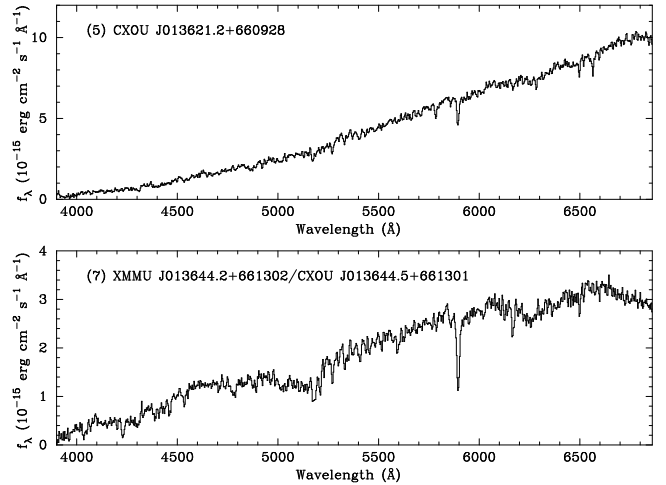}
\caption{Optical spectra of two of the X-ray sources in the {\em INTEGRAL}
error circle for IGR~J01363+6610.  The stars are listed as source \#5 
(top panel) and source \#7 (bottom panel) in Table~\ref{tab:sources}.
We identify source \#5 as a highly reddened G-type star and source \#7
as an early M-type star.\label{fig:stars}}
\end{figure}

We also obtained optical spectra for the only other two X-ray sources in the 
{\em INTEGRAL} error circle with optical/IR counterparts.  The spectra for
sources \#5 and \#7 are shown in Figure~\ref{fig:stars}, and we identify
them as G-type and early M-type stars, respectively.  Like Be X-ray binaries, 
LMXBs usually show H$\alpha$ in emission during outbursts \citep{fender09}
and during quiescence \citep{orosz02,cc06}.  Also, symbiotic systems often
show strong H$\alpha$ in emission \citep{cr97}.  However, there are 
counter-examples for both LMXBs and symbiotics.  In any case, as sources
\#5 and \#7 do not have emission lines, there is no reason to consider that 
they might be the IGR~J01363+6610 counterpart.

Thus, these observations show that the Be star that was previously considered
to be the likely counterpart of IGR~J01363+6610 is coincident with 
XMMU~J013549.5+661243, which was the brightest X-ray source in the {\em INTEGRAL} 
error circle during the 2009 {\em XMM-Newton} observation.  The fact that
we have now conclusively shown that the Be star is a transient X-ray source and 
that none of the other fainter X-ray sources in the {\em INTEGRAL} error circle 
have optical properties that would be expected of a hard X-ray source 
strengthens the association between the Be star and IGR~J01363+6610, and
we focus exclusively on the X-ray properties of the Be star XMMU~J013549.5+661243
in the remainder of this paper.

\subsection{XMMU~J013549.5+661243}

\begin{figure*}
%\plotone{fig4.ps}
\begin{center}
\includegraphics[clip,scale=0.9]{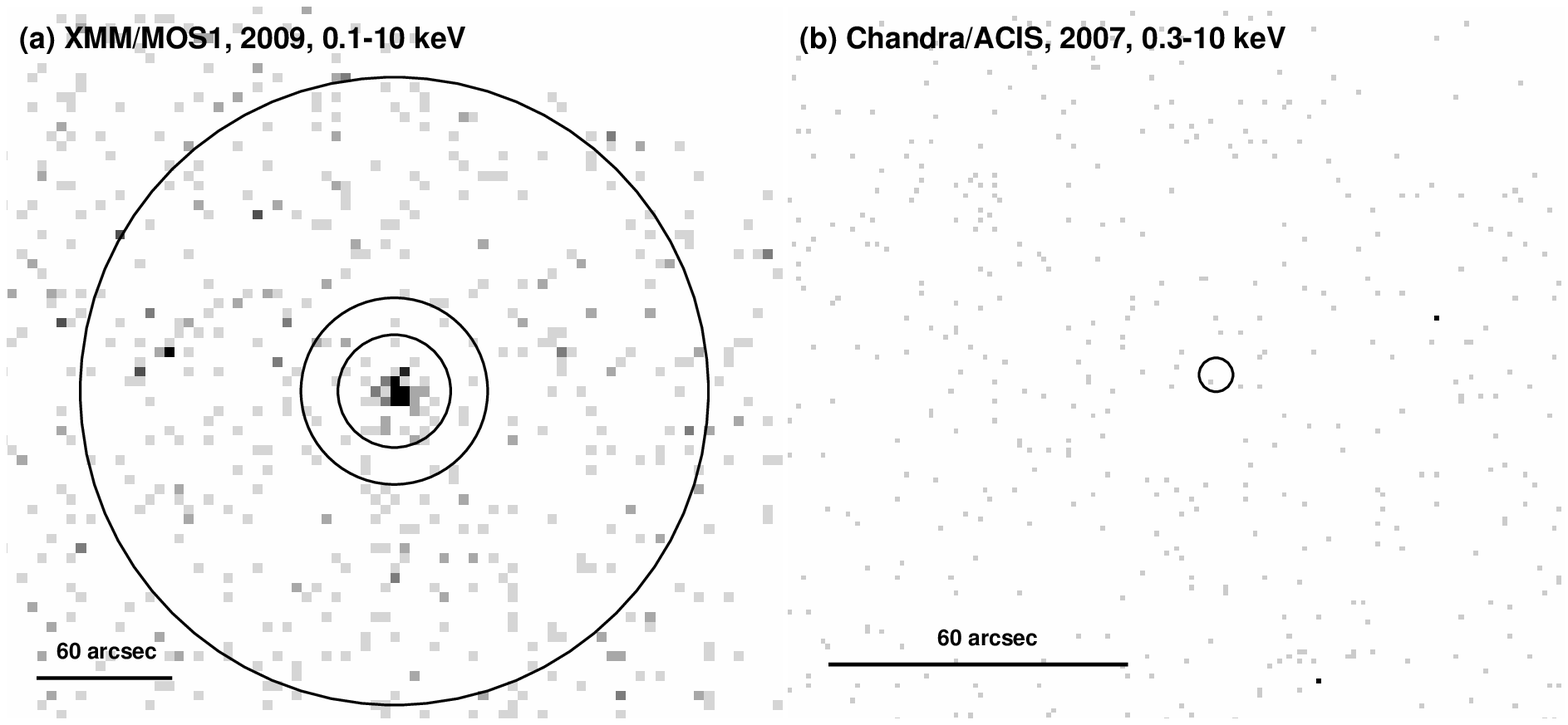}
\caption{{\em (a)} X-ray image from the 2009 July 31 {\em XMM-Newton}
observation showing the detection of XMMU~J013549.5+661243, which is
the very likely counterpart to IGR~J01363+6610.  The image uses
0.1--10~keV photons from the MOS1 instrument and was rebinned so that
each pixel is $4^{\prime\prime}$ wide.  The inner circle is the source 
region used for making the spectrum and the light curve.  The annulus 
(outer two circles) is the region used to estimate the background. 
{\em (b)} X-ray image from a 2007 {\em Chandra} observation.  The
image uses 0.3--10~keV photons from the ACIS instrument and was rebinned 
so that the pixels are $1^{\prime\prime}$.  The circle shows the 90\% 
confidence {\em XMM-Newton} error circle, highlighting the fact that 
IGR~J01363+6610 was not detected.  In both images, North is up and East 
is to the left.\label{fig:image}}
\end{center}
\end{figure*}

\subsubsection{Energy Spectrum}

We used the SAS tool {\ttfamily xmmselect} to produce MOS1, MOS2, and pn
energy spectra for XMMU~J013549.5+661243 and considered the recommendations 
given in a recent EPIC (MOS and pn) calibration document\footnote{The document
entitled ``EPIC status of calibration and data analysis'' by Guainazzi
et al.~(XMM-SOC-CAL-TN-0018) is based on results obtained using SAS-10.0.0.  
It was released on 2010 July 16 and can be found at
http://xmm2.esac.esa.int/external/xmm\_sw\_cal/calib/index.shtml.}
for event filtering and energy ranges.  For MOS, we used event filtering with 
the expression ``\#XMMEA\_EM \&\& PATTERN$<$=12'' and included events within 
an aperture with a $25^{\prime\prime}$ radius.  A background spectrum was 
extracted from an annulus centered on the source, and the source and background 
regions are shown in Figure~\ref{fig:image}a.  For the pn instrument, 
we used event filtering with the expression ``FLAG=0 \&\& PATTERN$<$=4'' 
and extracted spectra from source and background regions.  For all three 
instruments, we used the SAS tools {\ttfamily rmfgen} and {\ttfamily arfgen} 
to make response matrices.  For the MOS detectors, we used the 0.1--10~keV 
bandpass, and for the pn detector, we used the 0.2--12~keV bandpass.  We 
rebinned each spectrum, and we fitted the spectra using $\chi^{2}$ statistics.

\begin{figure}
%\plotone{fig5.ps}
\includegraphics[clip,scale=0.45]{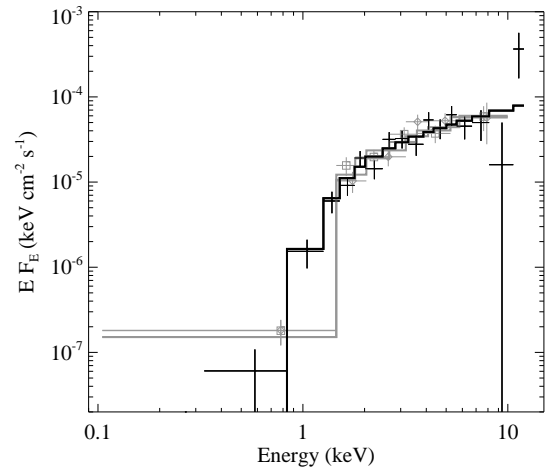}
\caption{{\em XMM-Newton} energy spectrum for XMMU~J013549.5+661243
fitted with an absorbed power-law model.  The model parameters are 
given in Table~\ref{tab:spec}, indicating a relatively hard spectrum 
($\Gamma = 1.4\pm 0.3$) but not a highly absorbed spectrum.  The
pn data are shown in black, and the MOS data are shown in grey with
diamonds plotted on MOS1 points and squares plotted on MOS2 points.
\label{fig:spec}}
\end{figure}

We used the XSPEC version 12 software for spectral fitting, and we 
tried several models.  The results of using an absorbed power-law
are shown in Table~\ref{tab:spec} and Figure~\ref{fig:spec}.  For
absorption, we used the photoelectric absorption cross sections
from \cite{bm92} and elemental abundances from \cite{wam00}, which 
correspond to the estimated abundances for the interstellar medium.
This model gives a good fit ($\chi^{2}/\nu = 23.1/26$), and requires
a relatively hard power-law photon index of $\Gamma = 1.4\pm 0.3$.  
The power-law fit gives a 0.2--12~keV unabsorbed flux of 
$(1.9^{+0.3}_{-0.2})\times 10^{-13}$ ergs~cm$^{-2}$~s$^{-1}$ and a column 
density of $N_{\rm H} = (1.5^{+0.7}_{-0.5})\times 10^{22}$~cm$^{-2}$, which 
is somewhat higher than the value through the Galaxy along the line of 
sight, $N_{\rm H} = 5.2\times 10^{21}$ cm$^{-2}$ \citep{kalberla05}.
Although this could indicate a small amount of absorption local to 
the source, it is much lower than the values near $10^{23}$ cm$^{-2}$
seen for the sources usually considered to be obscured HMXBs.  Another 
argument against local absorption of the X-ray source is that the value 
of $E(B-V) = 1.5$--2.0 corresponds to an $N_{\rm H}$ in the range 
(1.0--1.4)$\times 10^{22}$~cm$^{-2}$ \citep{ryter96}, which is consistent 
with the column density determined from the X-ray spectrum.

We also fitted the energy spectrum with blackbody and thermal 
Bremsstrahlung models, and the parameters are given in Table~\ref{tab:spec}.  
A fit with a 1.3~keV blackbody is similar in quality to the power-law 
($\chi^{2}/\nu = 21.5/26$), but the spectrum does not allow us to distinguish 
between thermal and non-thermal models.  However, at 2~kpc, the blackbody
model implies an emitting region with a radius of only $1.15\times 10^{3}$~cm, 
which is too small to be physical.  The Bremsstrahlung model requires a 
relatively high temperature ($>$11~keV), which makes it very similar to a 
power-law over the {\em XMM-Newton} bandpass.  Although we are not able to 
distinguish between models, it is important to note that the power-law fit 
above shows that the spectrum is hard, as expected for an IGR source and
especially for an HMXB.

\begin{table*}
\caption{Spectral Results for XMMU J013549.5+661243\label{tab:spec}}
\begin{minipage}{\linewidth}
\begin{center}
\footnotesize
\begin{tabular}{cc} \hline \hline
Parameter & Value\footnote{The errors on the parameters are for $\Delta\chi^{2} = 2.7$, 
corresponding to 90\% confidence for one parameter of interest.}\\ \hline
\multicolumn{2}{c}{Power-law model}\\ \hline
$N_{\rm H}$ & $(1.5^{+0.7}_{-0.5})\times 10^{22}$ cm$^{-2}$\\
$\Gamma$ & $1.4\pm 0.3$\\
$F_{PL}$ (unabsorbed, 0.2--12 keV) & $(1.9^{+0.3}_{-0.2})\times 10^{-13}$ ergs cm$^{-2}$ s$^{-1}$\\
$\chi^{2}/\nu$ & 23.1/26\\ \hline
\multicolumn{2}{c}{Blackbody model}\\ \hline
$N_{\rm H}$ & $(0.3^{+0.3}_{-0.2})\times 10^{22}$ cm$^{-2}$\\
$kT$ & $1.33^{+0.16}_{-0.15}$ keV\\
$R_{\rm km}^{2}/d_{10}^{2}$\footnote{The normalization for this model is parameterized in terms of the radius of a spherical emitting region in units of km ($R_{\rm km}$) and the distance in units of 10~kpc ($d_{10}$).} & $(3.3^{+1.6}_{-1.1})\times 10^{-3}$\\
$\chi^{2}/\nu$ & 21.5/26\\ \hline
\multicolumn{2}{c}{Bremsstrahlung model}\\ \hline
$N_{\rm H}$ & $(1.5^{+0.5}_{-0.4})\times 10^{22}$ cm$^{-2}$\\
$kT$ & $>$12 keV\\
$N_{bremss}$\footnote{The normalization for this model is $2.4\times 10^{-16} d^{-2} \int{n_{e} n_{i} {\rm d}V}$~cm$^{-5}$ where $d$ is the source distance, and $n_{e}$ and $n_{i}$ are, respectively, the electron and ion number densities within the volume $V$.} & $(2.9^{+1.5}_{-0.4})\times 10^{-5}$ cm$^{-5}$\\
$\chi^{2}/\nu$ &  22.6/26\\ \hline
\end{tabular}
\end{center}
\end{minipage}
\end{table*}

\subsubsection{Long- and Short-Term X-ray Variability}

Comparing the two panels of Figure~\ref{fig:image} shows the difference
between fluxes in 2009 (the {\em XMM-Newton} observation) and 2007 (the
{\em Chandra} observation).  We did not detect the Be star during a 5~ks 
{\em Chandra}/ACIS observation \citep{tomsick08a}, and 1 count is detected 
by ACIS within the {\em XMM-Newton} error circle.  Thus, using Poisson 
statistics, this corresponds to a 90\% confidence upper limit of $<$3.9 counts 
\citep{gehrels86}.  For the power-law spectral model measured by 
{\em XMM-Newton} ($\Gamma = 1.4$ and $N_{\rm H} = 1.5\times 10^{22}$~cm$^{-2}$), 
this corresponds to an absorbed 0.3--10~keV flux of $<$$1.6\times 10^{-14}$ 
ergs~cm$^{-2}$~s$^{-1}$ and an unabsorbed flux of $<$$2.9\times 10^{-14}$ 
ergs~cm$^{-2}$~s$^{-1}$, which is $>$6.6 times lower than the flux seen during 
the {\em XMM-Newton} observation.  During the original outburst detected by 
{\em INTEGRAL} in 2004, an 8--15~keV flux of 9 mcrab ($9.1\times 10^{-11}$ 
ergs~cm$^{-2}$~s$^{-1}$) was measured using the JEM-X instrument.  In the 
8--15~keV band, the flux measured with {\em XMM-Newton} in 2009 was 
$7.9\times 10^{-14}$ ergs~cm$^{-2}$~s$^{-1}$, which is a factor of 1,150 lower 
than the outburst level.  Thus, the quiescent upper limit obtained with 
{\em Chandra} represents a flux $>$7,600 times lower than the level seen 
during outburst.

\begin{figure}
%\plotone{fig6.ps}
\includegraphics[clip,scale=0.45]{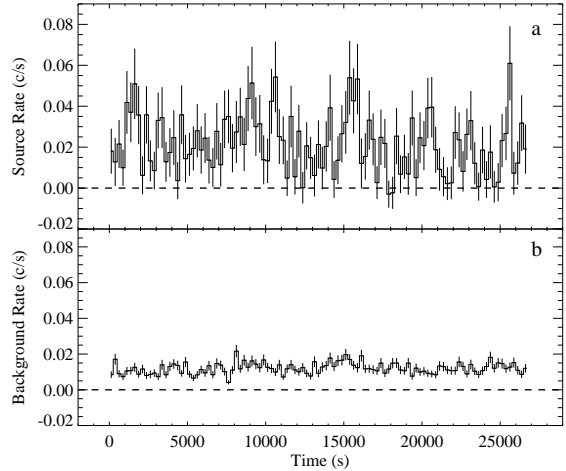}
\caption{Source {\em (a)} and background {\em (b)} {\em XMM-Newton}
light curves of XMMU~J013549.5+661243.  The rates are a combination 
of counts from all three instruments (pn, MOS1, and MOS2), and the
background rates have been normalized to the rates predicted in
the source region.  The energy range in both panels is 0.2--12~keV, 
and the time resolution is 250~s.  The source shows variability on 
this time scale.\label{fig:lc}}
\end{figure}

We also find evidence for shorter term variability during the {\em XMM-Newton}
observation.  Figure~\ref{fig:lc} shows the source and background rates
combined for all three {\em XMM-Newton} detectors in the 0.2--12~keV band
with 250~s time bins.  The light curve shows that the 0.2--12~keV rates can 
change from values as low as --$0.003\pm 0.007$ to $0.061\pm 0.018$.  We 
used a $\chi^{2}$ test to determine the significance of the variability.  
The best fit that can be obtained with a constant rate has a $\chi^{2} = 
142.5$ for 106 degrees of freedom, which indicates that the variability is 
significant at the 99\% confidence level.

\section{Discussion}

\subsection{The Nature of IGR~J01363+6610}

Our results show that XMMU~J013549.5+661243 is a hard X-ray source coincident 
with a Be star at a distance of $\sim$2~kpc.  At this distance, the inferred 
X-ray luminosity measured by {\em XMM-Newton} is $9.1\times 10^{31}$ ergs~s$^{-1}$.
While this would be a relatively high X-ray luminosity for an isolated B1 star, 
it is not too far above the values reported by \cite{ccm97} for such stars.  
However, in nearly all cases of isolated B-type stars, the X-ray emission is 
thermal with temperatures that are typically not above a few $\times 10^{6}$~K
\citep{ccm97,cohen00}, and their spectra fall very steeply above $\sim$0.25~keV.  
The very hard X-ray spectrum that we measure for XMMU~J013549.5+661243 is 
inconsistent with a thermal spectrum with such a low temperature, indicating 
that the Be star must have a compact binary companion that is emitting most 
of the X-ray emission.

While the X-ray luminosity and spectrum are not consistent with an isolated
star, the {\em XMM-Newton} data alone are not sufficient to determine the
nature of the compact binary companion.  There are known Be systems with
both white dwarf and neutron star companions that can produce hard spectra.
For example, there are several $\gamma$~Cas-like systems that may harbor
white dwarfs and have power-law spectral components with photon indices 
of 1.4--1.7 \citep{motch07}.  However, based on our study of the X-ray
sources in the IGR~J01363+6610 field, we argue that it is very likely 
that XMMU~J013549.5+661243 is the quiescent counterpart to IGR~J01363+6610.
In this case, the outburst detected by {\em INTEGRAL} indicates that the
source can produce an X-ray luminosity as high as $\sim$$10^{35}$ ergs~s$^{-1}$
along with variations in X-ray luminosity by a factor of $>$7,600 (when
the {\em Chandra} observation is considered as well).  In contrast, 
the $\gamma$~Cas-like systems have luminosities typically near 
$10^{33}$ ergs~s$^{-1}$, and while they show variability in flux by factors
of a few, they do not have large outbursts \citep{motch07}.  Thus, the
most likely interpretation of the available information is that
IGR~J01363+6610 is a Be X-ray binary with a neutron star (or perhaps a
black hole as discussed below) accretor.  In the following, we discuss
IGR~J01363+6610 in this context.

\subsection{Accretion Regimes and the X-ray Luminosity}

Models for X-ray emission from Be X-ray binaries are based on the picture of a 
neutron star with a relatively strong magnetic field ($B\sim 10^{12}$~G) and
a relatively slow rotation speed ($P_{\rm spin}\sim 0.1$--1,000~s) periodically 
accreting from the circumstellar disk around the Be star \citep{rv82,corbet86}.
In most systems, the orbits have non-zero eccentricity, and ``Type 1'' X-ray 
outbursts occur when the neutron star makes its closest approach to the star so 
that the outburst periodicity is equal to the orbital period, $P_{\rm orb}$.

Within this physical picture, there are two accretion regimes that are usually
considered.  One regime is at higher mass accretion rates when the accretion
pressure overcomes the centrifugal magnetic field barrier, and matter is accreted 
directly onto the poles of the neutron star \citep{corbet96}.  The second regime 
is during times when the accretion is inhibited by the magnetic field.  This is 
often called the propeller regime \citep{is75}, and the X-ray emission is much 
lower in this regime because the matter being accreted only reaches the neutron 
star magnetosphere rather than falling onto the neutron star surface.  A gap is 
predicted between the minimum luminosity in which the system can be in the
direct accretion regime and the maximum luminosity in which the system can 
be in the propeller regime \citep{corbet96}, and we define $\Delta$ to be 
equal to this luminosity ratio.  For typical neutron star properties (1.4~\Msun~mass, 
10~km radius, $10^{12}$~G magnetic field), $\Delta = 170 (P_{\rm spin}/1~{\rm s})^{2/3}$ 
\citep{campana02a}, and values of $\Delta$ range from $\sim$100 for fast rotators 
to $\sim$10,000 for slow rotators \citep{corbet96}.  

Assuming a distance of 2~kpc and a spectral shape consistent with the power-law
model measured using {\em XMM-Newton}, IGR~J01363+6610 has been observed by 
{\em INTEGRAL} at $L_{\rm x} = 1.04\times 10^{35}$ ergs~s$^{-1}$ in 2004, by
{\em Chandra} at $<$$1.4\times 10^{31}$ ergs~s$^{-1}$ in 2007, and by {\em XMM-Newton}
at $9.1\times 10^{31}$ ergs~s$^{-1}$ in 2009.  These luminosities are unabsorbed
values measured in or extrapolated into the 0.2--12~keV band.  The 2004 outburst
luminosity is at the lower end of typical values for normal (``type I'') Be X-ray 
binary outbursts, which are in the $10^{35}$ to $10^{37}$ ergs~s$^{-1}$ range
\citep{swr86,wilson05}.  Still, it is likely (but perhaps not certain if the 
neutron star spin or magnetic field strength have extreme values) that the source 
reached a high enough accretion rate to enter the direct accretion regime.  
Following \cite{corbet96} and \cite{campana02a}, we derive an upper limit on
the luminosity produced by magnetospheric emission of 
$L_{\rm m} < 1.04\times 10^{35}/\Delta = 6.1\times 10^{32} (P_{\rm spin}/1~{\rm s})^{-2/3}$.
Thus, under the assumptions that the neutron star properties (mass, radius, 
and magnetic field strength) are typical and the system reached the direct 
accretion regime during the 2004 outburst, the luminosity measured by 
{\em XMM-Newton} in 2009 implies a neutron star spin period less than 17~s 
for IGR~J01363+6610.

While the propeller regime may extend to very low luminosities, a third 
regime to consider is when the mass accretion rate onto the neutron star
drops to zero.  In this case, any X-ray emission produced could have
contributions from the Be star and from the neutron star.  Emission
from the neutron star would be thermal in origin and would be extremely
soft.  Such components are seen at temperatures near 50--150~eV in LMXBs
\citep{bbr98,campana98,degenaar09} and although it is unclear whether 
this regime occurs for Be X-ray binaries, it has probably been seen 
for the HMXB Supergiant Fast X-ray Transient (SFXT) IGR~J17544--2619 
\citep{intzand05}.  During the {\em XMM-Newton} observation of 
IGR~J01363+6610, we are not seeing this regime since we measure a hard
spectrum, which is either non-thermal or has a much higher temperature
($kT = 1.3$~keV) than would be expected from a non-accreting neutron 
star.  Since we did not detect the source during the {\em Chandra} 
observation, the spectrum is not constrained.  Assuming the same $N_{\rm H}$ 
as seen during outburst ($N_{\rm H} = 1.6\times 10^{22}$ cm$^{-2}$), we can 
put limits on the flux and the temperature of the neutron star by requiring 
an ACIS count rate below the upper limit given above.  We use the magnetic 
neutron star model {\it nsa} \citep{Pavlov95}, assuming emission 
from the entire neutron star, a mass of 1.4\Msun, a radius of 10 km, a 
distance of 2 kpc, and a magnetic field of $B = 10^{12}$~G.  The upper 
limit on the (unredshifted) surface temperature is then 80~eV, and the 
unabsorbed 0.2--12~keV luminosity upper limit is $<$$2.4\times 10^{32}$ 
ergs~s$^{-1}$.  Changes in the assumptions will affect this, particularly 
changes in the assumed $N_{\rm H}$ or distance.  For instance, a choice of 
$N_{\rm H} = 2\times 10^{22}$ cm$^{-2}$ gives $kT < 85$~eV and 
$L_{\rm x} < 3.2\times 10^{32}$ ergs~s$^{-1}$, or a distance of 3~kpc gives
$kT < 88$~eV and $L_{\rm x} < 3.7\times 10^{32}$ ergs~s$^{-1}$.  If the X-ray 
emission in this quiescent state is dominated by hot spots at the polar 
caps, their temperature may be higher, but the temperature of the rest 
of the neutron star surface must be even lower.

\subsection{The Quiescent Luminosity and Possible Implications}

The above analysis suggests that the IGR~J01363+6610 luminosities can be 
explained within the standard picture for Be-neutron star X-ray binaries, 
but the luminosity upper limit that we infer from the {\em Chandra} observation 
is lower than has been previously reported from these systems.  \cite{campana02a}
present a study that focuses on quiescent X-ray observations of 3 Be X-ray
binaries, and their luminosities are (1--3)$\times 10^{35}$ ergs~s$^{-1}$ 
(0.1--10~keV) for A~0538--66, (0.8--2)$\times 10^{33}$ ergs~s$^{-1}$ 
(0.5--10~keV) for 4U~0115+63, and $5\times 10^{32}$ ergs~s$^{-1}$ (0.5--10~keV)
for V~0332+53.  In addition, quiescent luminosities of (3--9)$\times 10^{33}$ 
ergs~s$^{-1}$ and (2--4.5)$\times 10^{33}$ ergs~s$^{-1}$ have been reported for 
GRO~J2058+42 and A~0535+26, respectively \citep{wilson05,negueruela00}.
The candidate Be X-ray binary XTE~J1829--098 has been observed at a luminosity 
of $3\times 10^{32}$ ($d/10$~kpc)$^{2}$ ergs~s$^{-1}$ \citep{hg07}, which is 
low for a Be X-ray binary, but still not as low as we find for IGR~J01363+6610.
It should, however, be noted that, for A~0535+26, \cite{negueruela00} found 
the source at this luminosity after it had previously been reported to
have a quiescent luminosity two orders of magnitude higher \citep{motch91}.
Thus, it is clear that even when they are not in outburst, these sources 
exhibit a large amount of variability, and many observations may be required
for each source to define a quiescent luminosity.  

While the {\em Chandra} upper limit of $<$$1.4\times 10^{31}$ ($d/2$~kpc)$^{2}$ 
ergs~s$^{-1}$ obtained with a hard spectrum as seen in outburst
is significantly lower than has been found for other Be X-ray 
binaries, this conclusion does depend on the validity of the distance 
determination.  The 2~kpc distance from \cite{reig05} depends on the 
measurement of the $V$-band magnitude, which is given in \cite{reig05}
to high precision ($V = 13.29\pm 0.02$, based on three measurements), the 
extinction, and the spectral type.  \cite{reig05} give a value of 
$E(B-V) = 1.6$, and our determination is consistent with this suggesting
that the uncertainty in the distance due to the measurement of extinction
is not large.  Probably the largest uncertainty is related to the spectral
type.  \cite{reig05} use a spectral type of B1V and an absolute magnitude
of $M_{V}$ = --3.2 for the 2~kpc estimate, but they indicate that the star 
could be a more luminous sub-giant (i.e., B1IV). The difference in the 
absolute magnitude of a main sequence B-type star compared to a B-type 
sub-giant is 0.6--0.7 magnitudes \citep{cox00}, so it is unlikely that the 
star is more luminous than $M_{V}\sim$ --4.0, which corresponds to a distance
upper limit near 3~kpc.  This, in turn, indicates an upper limit on 
the quiescent X-ray luminosity of $<$$3.2\times 10^{31}$ ergs~s$^{-1}$, 
which is still significantly lower than the X-ray luminosities given
above for comparison.

Although it is not clear why the quiescent luminosity is so much lower
for IGR~J01363+6610, there are several potentially interesting possibilities.
Among the more trivial is the possibility that the short (5~ks) 
{\em Chandra} observation occurred during an X-ray eclipse.  Although
eclipses are not common for Be X-ray binaries due to their typically
wide orbits, we currently cannot rule out this possibility since the
orbit and binary inclination for IGR~J01363+6610 are unconstrained.
It is also possible that {\em Chandra} did not detect the source either 
because the compact object was highly obscured during the {\em Chandra} 
observation or that {\em Chandra} happened to catch the source during a 
low point in its normal short-term variability. Finally, there is also 
the possibility that the circumstellar disk, which was strongly present 
in 2004 and 2009 based on the detection of the H$\alpha$ emission line 
with an EW of $\sim$--50\AA, had dissipated during the 2007 {\em Chandra}
observation.  Although dissipation and re-formation of a circumstellar 
disk on this time scale did occur in the case of GRO~J1008--57 
\citep{coe07}, and significant disk loss was observed for IGR~J06074+2205 
on a time scale of $\sim$3 years \citep{rzg10}, in both of these cases, 
the H$\alpha$ lines were significantly weaker than measured for 
IGR~J01363+6610.  There is no precedent that we are aware of for 
dissipation and re-formation of a circumstellar disk in a system 
with an H$\alpha$ EW as large as the one measured for IGR~J01363+6610.  
While there are possible explanations for the luminosity upper limit 
measured over a short period of time by {\em Chandra}, these explanations 
do not apply to the {\em XMM-Newton} observation.  The luminosity 
measured by {\em XMM-Newton} would also make it one of, if not the, 
lowest luminosity Be X-ray binaries.  Thus, it is likely that the low 
luminosities measured for IGR~J01363+6610 require that this system has 
some unusual physical properties.

The Be X-ray binaries with known compact object type have neutron stars,
but binary evolution models predict that between zero and two of the 64 
known Be X-ray binary systems harbor a black hole instead of a neutron star 
\citep{bz09}.  The observational signatures for Be X-ray binaries with
black holes are not as clear as neutron star signatures, such as pulsations.
At high mass accretion rates, one would expect the X-ray emission from
black holes to be softer and brighter due to the presence of an inner
accretion disk that cannot form in systems with highly magnetized neutron
stars due to the magnetosphere.  At moderate to low mass accretion rates, 
accreting black holes typically have hard spectra \citep{rm06}.  At very 
low mass accretion rates, black hole systems would likely be fainter than 
neutron star systems because the latter exhibit magnetospheric and surface
emission, which are both absent from black holes.  The expected hard 
spectrum and the faint quiescent luminosity are both consistent with the 
observed properties of IGR~J01363+6610.

There may, however, be a less exotic explanation if the binary orbit
of IGR~J01363+6610 causes a neutron star to sample an unusually 
low density part of the circumstellar disk.  This could occur if the 
system has a long orbital period with relatively low eccentricity.
As pointed out by \cite{reig05}, the EW of the H$\alpha$ line for 
IGR~J01363+6610 is one of the strongest of any Be X-ray binary, 
suggesting the presence of a large circumstellar disk, which might
favor a large and nearly circular orbit over a highly eccentric orbit.  
While the system may have a large circumstellar disk, \cite{no01} have
shown that tidal interations of the neutron star can lead to truncation
of the circumstellar disk in low eccentricity systems.  In addition
to providing a possible explanation for the low quiescent luminosity
of IGR~J01363+6610, the fact that the disk can be truncated to a size
smaller than the Be star's Roche lobe means that the system will not
show luminous outbursts \citep{on01}, which is consistent with what 
has been seen so far for IGR~J01363+6610.  

Like IGR~J01363+6610, the orbital parameters for the other Be X-ray
binaries with the strongest H$\alpha$ emission lines, A~1118--616 
\citep{coe94} and IGR~J01583+6713 \citep{kaur08}, are not known.
It has been suggested that A~1118--616 has a nearly circular orbit 
based on its having a small number of outbursts with no clear 
periodicity \citep{coe94}.  It has been argued that both of these
systems have relatively long orbital periods ($\sim$0.6--2.2 years) 
based on measured neutron star spin periods of 405.6~s and 469.2~s
and the $P_{\rm orb}$--$P_{\rm spin}$ relationship for Be X-ray binaries 
\citep{corbet86,coe94,kaur08,doroshenko10}.  However, for 
IGR~J01583+6713, the detection of X-ray pulsations was noted as being 
marginal in \cite{kaur08}.  Although there is no direct indication 
of the orbital or spin period for IGR~J01363+6610, it should be noted 
that the limit of $P_{\rm spin} < 17$~s mentioned above could be taken 
as evidence for a smaller $P_{\rm orb}$; however, we view the evidence 
for a limit on the spin period as being relatively weak.  

\section{Summary and Conclusions}

In summary, the detection of the variable hard X-ray source 
XMMU~J013549.5+661243 at the location of a Be star provides confirmation 
that IGR~J01363+6610 is a Be X-ray binary.  Although some of its 
properties suggest that it is a fairly typical HMXB, its 2007 
non-detection by {\em Chandra} indicate a quiescent X-ray luminosity 
that is significantly lower than has been measured previously for other 
Be X-ray binaries.  While some possible explanations for the low 
luminosity (an eclipse, a large change in the column density or spectrum, 
an extended drop in mass accretion rate, or dissipation of the Be star's 
circumstellar disk) may be consistent with relatively normal Be X-ray 
binary properties, other possibilities would require that IGR~J01363+6610 
has unusual properties.  One possibility is that the system has a large 
and relatively circular orbit, which could explain the low duty cycle 
for outbursts and, perhaps, the low quiescent luminosity.  Another very 
interesting possibility is that this could be a Be-black hole system.  
The former suggestion could be confirmed by a measurement of the orbital 
period (although this will be challenging since the source does not 
seem to produce regular outbursts), and the latter suggestion could 
be refuted with the detection of pulsations during another outburst from 
the source.

\acknowledgments

JAT acknowledges partial support from NASA {\em XMM-Newton} Guest Observer award number 
NNX09AP91G.  COH acknowledges support from an NSERC Discovery Grant.  We thank Ms.~Jia 
Liu for assistance with the optical observations and spectral reductions.  We acknowledge
helpful comments from the referee, Ignacio Negueruela.  This publication makes use of 
data products from the Two Micron All Sky Survey, which is a joint project of the 
University of Massachusetts and the Infrared Processing and Analysis Center/California 
Institute of Technology, funded by NASA and the National Science Foundation.  This 
research makes use of the USNOFS Image and Catalog Archive operated by the United States 
Naval Observatory, Flagstaff Station and the SIMBAD database, operated at CDS, Strasbourg, 
France.

% BIBLIOGRAPHY
%\bibliographystyle{jwapjbib}
%\bibliography{refs}

\end{document}